\documentclass[conference]{IEEEtran}
\usepackage{amsmath,amsfonts}
\usepackage{algorithm}
\usepackage{algorithmic}
\usepackage{array}
\usepackage[caption=false,font=normalsize,labelfont=sf,textfont=sf]{subfig}
\usepackage{graphicx}
\usepackage{multicol,multienum}
\usepackage{multirow}
\usepackage{textcomp}
\usepackage{stfloats}
\usepackage{url}
\usepackage{verbatim}
\usepackage{bm}
\usepackage{xcolor}
\usepackage{makecell}
\def\BibTeX{{\rm B\kern-.05em{\sc i\kern-.025em b}\kern-.08em
		T\kern-.1667em\lower.7ex\hbox{E}\kern-.125emX}}
\usepackage{balance}
\usepackage{geometry} 
\begin{document}

\title{MIMO-OTFS-Based Semantic Communication for High-Mobility Scenarios}

\author{
\IEEEauthorblockN{Yu Zhang$^1$, Jiarui Yan$^1$, Yue Liu$^2$, Tenglun Ke$^3$, Yimeng Wang$^3$, and Zhijin Qin$^1$} \\
\IEEEauthorblockA{$^1$Department of Electronic Engineering, Tsinghua University, Beijing 100084, China \\
$^2$Faculty of Applied Sciences, Macao Polytechnic University, MacaoSAR, China\\
$^3$Beijing Institute of Remote Sensing Equipment, Beijing 100854, China\\
Email: zhangyu06@mail.tsinghua.edu.cn}
}
\maketitle

\begin{abstract}
In high-mobility scenarios with time-frequency doubly-selective channels, existing semantic communication systems suffer significant performance degradation. To address this issue, we propose a semantic communication framework that synergistically integrates multiple-input multiple-output orthogonal time frequency space (MIMO-OTFS) with semantic-aware sub-channel allocation. First, an entropy module is employed to evaluate importance of different semantic features, and the Kendall correlation coefficient is used to quantify the alignment between semantic importance and sub-channel conditions. Subsequently, joint optimization of the encoder and decoder is achieved through a comprehensive loss function that balances image classification accuracy, reconstruction quality, and sub-channel matching degree. Experimental results confirm the superior reconstruction quality of our proposed framework compared to conventional semantic communication systems based on orthogonal frequency division multiplexing in high-mobility channel environment.
\end{abstract}

\begin{IEEEkeywords}
Semantic communication, MIMO-OTFS, semantic importance, sub-channel allocation
\end{IEEEkeywords}

\section{Introduction}
The advent of 6G and beyond wireless networks is driving the evolution of communication paradigms from merely focusing on bit-level accuracy towards the goal of successful task-oriented semantic transmission \cite{Qin2023AGS}. Semantic communication embodies a paradigm shift wherein the focus evolves from bit-level accuracy to conveying the underlying meaning and intent of information. Through task-oriented information extraction at the transmitter and intelligent semantic interpretation at the receiver, semantic communication can significantly reduce redundant transmission and enhance transmission efficiency. These advantages are particularly critical in high-mobility scenarios, such as high-speed railways, autonomous vehicles, and low-altitude aviation. 

There has been surging interest in semantic communication from both academia and industry over the past few years. Early studies \cite{Bourtsoulatze2019DeepJS, Xie2021DeepLE, Weng2021SemanticCS, Tung2022DeepWiVe} developed joint semantic-channel coding (JSCC) that outperforms traditional separated coding under low signal-to-noise-ratio (SNR) conditions. However, incompatibility of JSCC with existing digital systems hindered its practical deployment. To bridge this gap, Bo \emph{et al.} \cite{Bo2024JointCM} developed a joint coding and modulation (JCM) framework which converts source data into discrete constellation symbols through probability converters, achieving joint optimization of coding and modulation. Zhang \emph{et al.} \cite{Zhang2025FromAT} further proposed a multi-stage JCM method and modelled the modulation as a constrained quantization process, which is approximated by using scaling factors and artificial noise.

Under multipath fading channels, both JSCC and JCM-based semantic communication systems suffer from serious inter-symbol interference. To mitigate this issue, researchers have introduced orthogonal frequency division multiplexing (OFDM) technology with excellent resistance into semantic communication systems. Yang \emph{et al.} \cite{Yang2022OFDMGD} incorporated multipath fading channels and OFDM modulation as differentiable layers into a JSCC-based end-to-end learning framework. Jiang \emph{et al.} \cite{Jiang2023WirelessSC} integrated constellation modulation and OFDM modules in the JSCC framework, improving semantic recovery ability under different channel conditions. Liu \emph{et al.} \cite{Liu2024OFDMBD} designed task-related semantic importance metrics and corresponding subcarrier allocation strategies, further boosting transmission performance over multipath fading channels.

Nevertheless, in high-mobility scenarios characterized by pronounced Doppler shift and time-frequency doubly-selective fading, existing semantic communication systems still face considerable performance degradation. Addressing the challenges posed by high-mobility scenarios remains a critical issue. Orthogonal time frequency space (OTFS) modulation offers a compelling alternative \cite{Liu2025OTFSVO}. Its key advantage stems from delay-Doppler (DD) domain processing, which effectively mitigates time-varying channel effects by rendering them nearly stationary. When combined with multiple-input multiple-output (MIMO) technology, MIMO-OTFS provides substantial diversity and multiplexing gains, establishing a highly reliable physical-layer foundation \cite{Keskin2024IntegratedSA}.

More importantly, most existing semantic communication systems treat semantic modeling and physical-layer transmission as separate entities, lacking deep integration. A truly intelligent semantic communication architecture should facilitate dynamic and synergistic interaction between the semantic and physical layers. Specifically, such a system should be capable of evaluating the relative importance of different semantic components and adaptively mapping them to available physical resources. 
Inspired by this, we present a MIMO-OTFS-based semantic communication framework, referred to as MIMO-OTFS-SC, which deeply integrates semantic importance assessment with sub-channel allocation. Our main contributions are twofold:

\begin{itemize}
  \item We develop a novel MIMO-OTFS-SC system architecture specifically designed for reliable transmission in doubly-selective channel environments. This architecture effectively converts time-varying multipath channels into near-stationary delay-Doppler domain representations through innovative signal modulation techniques, providing a robust physical-layer foundation for semantic communication in high-mobility scenarios.
  \item We design a semantic-aware sub-channel allocation (SA-SCA) strategy that employs an entropy model to evaluate feature significance and utilizes Kendall correlation optimization to dynamically align critical semantic features with high-quality sub-channels. This strategy ensures optimal matching between semantic importance and channel conditions, enabling intelligent resource allocation.
\end{itemize}

Simulation results confirm that our proposed framework outperforms conventional OFDM-based semantic communication systems in terms of image reconstruction quality and classification accuracy under doubly-selective channels.

Following this introduction, Section II describes the MIMO-OTFS-SC system architecture. Section III elaborates on the SA-SCA scheme. Section IV discusses simulation results and performance evaluation. Conclusion is given in Section V. 

\emph{Notations}: Vectors are indicated by bold lowercase ${\mathbf{a}}$, and matrices by bold uppercase letters ${\mathbf{A}}$. ${{\mathbf{I}}_N}$ is an ${N \times N}$ identity matrix. ${\mathbb{R}^{M \times N}}$ means an ${M \times N}$ real matrix and ${\mathbb{C}^{M \times N}}$ means an ${M \times N}$ complex matrix. ${{\mathbf{A}}^T}$ and ${{\mathbf{A}}^H}$ represent the transpose and conjugate transpose of $\mathbf{A}$, respectively. ${{{\bf{F}}_M}}$ denotes the $M$-point fast Fourier transform (FFT) matrix. Horizontal and vertical concatenation of $\bf{a}$ and $\bf{b}$ are represented by $\left[ {{\bf{a}},{\bf{b}}} \right]$ and $\left[ {{\bf{a}};{\bf{b}}} \right]$. $\mathbf{A} \otimes \mathbf{B}$ means Kronecker product of $\mathbf{A}$ and $\mathbf{B}$. $\mathbb{E}\left\{ {\mathrm{\cdot}} \right\}$ denotes expectation. $\mathcal{CN}\left( {{\mathbf{a}},{\mathbf{A}}} \right)$ indicates a complex Gaussian vector characterized by mean ${\mathbf{a}}$ and covariance ${\mathbf{A}}$.

\section{Framework of the MIMO-OTFS-SC System}


\begin{figure} [bt!]
	\centering
	\subfloat[]{
		\includegraphics[width=3.4in]{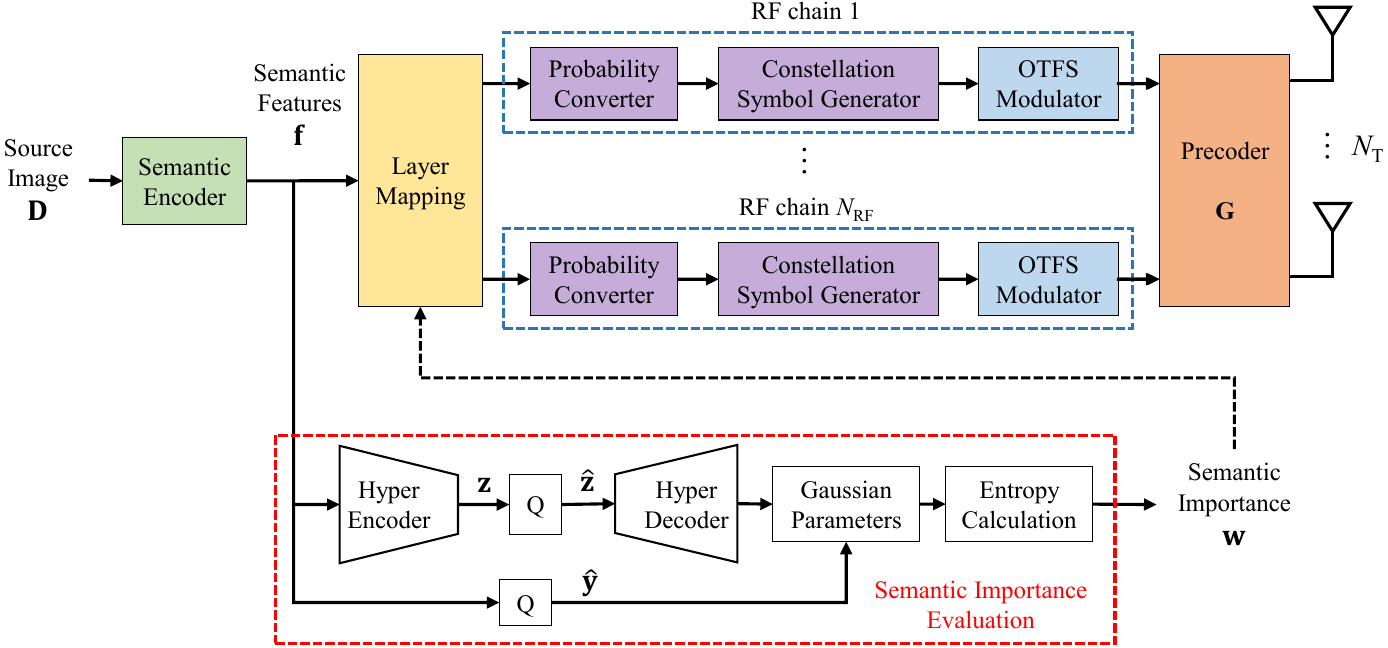}} \\
	\subfloat[]{
		\includegraphics[width=3.4in]{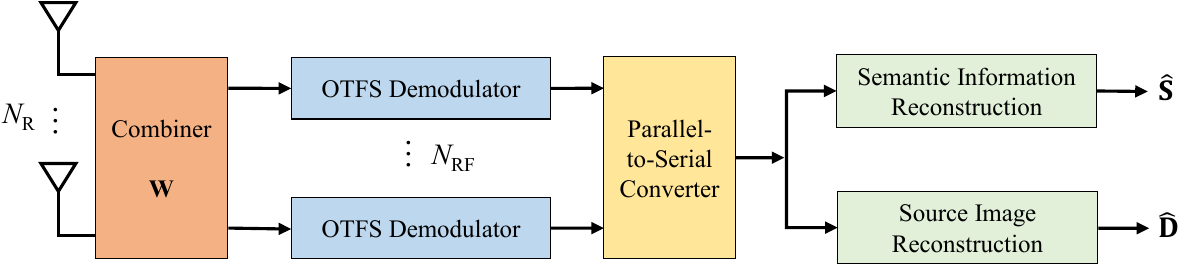}}
	\caption{The proposed MIMO-OTFS-SC system architecture. (a) Transmitter, (b) Receiver.}
	\label{fig:system-model}
\end{figure}

Fig. \ref{fig:system-model} depicts the architecture of the proposed MIMO-OTFS-SC system. At the transmitter side, $N_{\rm{T}}$ antennas are used, while the receiver is equipped with $N_{\rm{R}}$ antennas, both featuring $N_{\rm{RF}}$ radio-frequency (RF) chains for image transmission.

At the TX, a neural network is employed to extract semantic features $\bf{f}={\mathcal{F}}_{\bm{\alpha}}(\bf{D})$ from the source data ${\bf{D}}$, where $\mathcal{F}_{\bm{\alpha}}$ denotes the semantic encoder parameterized by trainable weights $\bm{\alpha}$. Let $\bf{S}$ represent the inherent semantic information in the source data. The semantic encoder outputs a feature sequence ${\bf{f}}=[f_1,f_2,...,f_K]^T$, where $K$ denotes the total number of semantic features.

The semantic features are processed through an entropy model (EM) to obtain the semantic importance vector ${\bf{w}}$. The specific details of the EM will be described in Section III. Based on their semantic importance and the quality of MIMO-OTFS sub-channels, the semantic feature sequence ${\bf{f}}$ are mapped to $N_{\rm{RF}}$ RF chains, forming a semantic feature matrix ${\bf{E}} = \left[ {{{\bf{e}}_1},{{\bf{e}}_2}...,{{\bf{e}}_{{N_{{\rm{RF}}}}}}} \right] \in \mathbb{R}{^{{N_{\rm{S}}} \times {N_{{\rm{RF}}}}}}$, where ${{\bf{e}}_{{n}}} \in {^{{N_{\rm{S}}} \times 1}}, n=1,2,...,N_{\rm{RF}}$ denotes the semantic feature vector assigned to the $n$-th RF chain, and $N_{\rm{S}} = K/N_{\rm{RF}}$.

Each RF chain independently processes its assigned semantic features through constellation modulation and OTFS modulation. For the $n$-th RF chain, $\bf{e}_n$ is first fed into a probability converter to obtain the transition probability from semantic features to constellation symbols. Based on the transition probability, a constellation symbol generator then produces the corresponding constellation symbols ${\bf{M}}_n$.

The symbols ${\bf{M}}_n$ is mapped to the DD-domain, forming OTFS frames denoted as ${\bf{X}}_n \in \mathbb{C}^{P \times M \times N}$, where $P$, $M$, and $N$ are the number of OTFS frames, the number of symbols in the delay dimension, and the number of symbols in the Doppler dimension, respectively. For each OTFS frame ${\bf{X}}_{n,p} \in \mathbb{C}^{M \times N}, p=1,2,...,P$, we apply the inverse symplectic fast Fourier transform (ISFFT) followed by the Heisenberg transform
\begin{equation}\label{}
  {{\bf{S}}_{n,p}} = \underbrace {{\bf{F}}_M^H\left( {\underbrace {{{\bf{F}}_M}{{\bf{X}}_{n,p}}{\bf{F}}_N^H}_{{\rm{ISFFT}}}} \right)}_{{\rm{Heisenberg\; transform}}} = {{\bf{X}}_{n,p}}{\bf{F}}_N^H ,
\end{equation}
where ${{\bf{S}}_{n,p}} \in \mathbb{C}^{M \times N}$ is the time-domain transmit signal of the $p$-th frame of the $n$-th RF chain.

By vectorizing ${{\bf{S}}_{n,p}}$, we obtain
\begin{equation}\label{}
  {{\bf{s}}_{n,p}} = {\rm{vec}}\left( {{{\bf{S}}_{n,p}}} \right) = \left( {{\bf{F}}_N^H \otimes {{\bf{{\bf I}}}_M}} \right){{\bf{x}}_{n,p}} \in \mathbb{C}^{MN \times 1},
\end{equation}
where ${{\bf{x}}_{n,p}}={\rm{vec}}({{\bf{X}}_{n,p}}) \in \mathbb{C}^{MN \times 1}$. Assuming a time slot is composed of one OTFS frame, the time-domain signals from all $N_{\rm{RF}}$ RF chains for the $p$-th time slot are then combined, resulting in
\begin{equation}\label{}
  {{\bf{s}}_p} = \left( {{{\bf{{I}}}_{{N_{{\rm{RF}}}}}} \otimes \left( {{\bf{F}}_N^H \otimes {{\bf{{I}}}_M}} \right)} \right){{\bf{x}}_p} ,
\end{equation}
where ${{\bf{s}}_p} = \left[ {{{\bf{s}}_{1,p}};{{\bf{s}}_{2,p}};...;{{\bf{s}}_{{N_{{\rm{RF}}}},p}}} \right] \in \mathbb{C}^{N_{\rm{RF}}MN \times 1}$ and ${{\bf{x}}_p} = \left[ {{{\bf{x}}_{1,p}};{{\bf{x}}_{2,p}};...;{{\bf{x}}_{{N_{{\rm{RF}}}},p}}} \right] \in \mathbb{C}^{N_{\rm{RF}}MN \times 1}$. Through an $N_{\rm{T}}MN \times N_{\rm{RF}}MN$ precoder ${\bf{G}}_p$, the signal is mapped onto the antennas for transmission, yielding ${\bf{y}}_p={\bf{G}}_p{\bf{s}}_p \in \mathbb{C}^{N_{\rm{T}}MN \times 1}$.

At the RX, the time-domain received signal is given by
\begin{equation}\label{}
  {\bf{r}}_p = {\bf{H}}_p{\bf{y}}_p + {\bf{n}}_p \in \mathbb{C}^{N_{\rm{R}}MN \times 1},
\end{equation}
where ${\bf{H}}_p \in \mathbb{C}^{N_{\rm{R}}MN \times N_{\rm{T}}MN}$ represents the time-domain channel matrix, and ${{\bf{n}}_p} \sim {\cal C}{\cal N}\left( {{\bf{0}},\sigma _p^2{{\bf{{I}}}_{{N_{\rm{R}}}MN}}} \right) \in \mathbb{C}^{N_{\rm{R}}MN \times 1}$ denotes the complex additive white Gaussian noise with $\sigma _p^2$ being the noise power.

The processing procedure at the RX is the reverse of that at the TX. The received signal is first processed by an $N_{\rm{R}}MN \times N_{\rm{RF}}MN$ combiner ${\bf{W}}_p$, yielding
\begin{equation}\label{}
  {{{\bf{\hat s}}}_p} = {\bf{W}}_p^H{{\bf{r}}_p} = {\bf{W}}_p^H{{\bf{H}}_p}{{\bf{G}}_p}{{\bf{s}}_p} + {\bf{W}}_p^H{{\bf{n}}_p} .
\end{equation}

The vector ${{{\bf{\hat s}}}_p}$ is then distributed across $N_{\rm{RF}}$ RF chains, each of which sequentially applies the Wigner transform and the symplectic fast Fourier transform (SFFT) to obtain the DD-domain signal. The resulting DD-domain received symbol vector ${{{\bf{\hat x}}}_p} \in \mathbb{C}^{N_{\rm{RF}}MN \times 1}$ is expressed as
\begin{equation}\label{equ:input_output}
  \begin{array}{l}
{{{\bf{\hat x}}}_p} = \left( {{{\bf{{I}}}_{{N_{{\rm{RF}}}}}} \otimes \left( {{{\bf{F}}_N} \otimes {{\bf{{I}}}_M}} \right)} \right){{{\bf{\hat s}}}_p}\\
 = \underbrace {\left( {{{\bf{{I}}}_{{N_{{\rm{RF}}}}}} \otimes \left( {{{\bf{F}}_N} \otimes {{\bf{{I}}}_M}} \right)} \right){\bf{W}}_p^H{{\bf{H}}_p}{{\bf{G}}_p}\left( {{{\bf{I}}_{{N_{{\rm{RF}}}}}} \otimes \left( {{\bf{F}}_N^H \otimes {{\bf{I}}_M}} \right)} \right)}_{{\rm{DD}} - {\rm{domain\; effective\; channel}}}{{\bf{x}}_p} \\
  + \underbrace {\left( {{{\bf{{I}}}_{{N_{{\rm{RF}}}}}} \otimes \left( {{{\bf{F}}_N} \otimes {{\bf{{I}}}_M}} \right)} \right){\bf{W}}_p^H{{\bf{n}}_p}}_{{\rm{DD}} - {\rm{domain\; effective\; noise}}} \;.
\end{array}
\end{equation}

By combining all $P$ DD-domain received frames, we obtain the recovered semantic feature vector ${\bf{\hat x}} = \left[ {{{{\bf{\hat x}}}_1};{{{\bf{\hat x}}}_2};...;{{{\bf{\hat x}}}_P}} \right]$, which is used for semantic information reconstruction and source image reconstruction
\begin{equation}\label{}
  {\bf{\hat S}} = {{\cal{G}}_{\bm{\beta }}}\left( {{\bf{\hat x}}} \right), \;\; {\bf{\hat D}} = {{\cal{G}}_{\bm{\delta }}}\left( {{\bf{\hat x}}} \right) ,
\end{equation}
where ${{\cal{G}}_{\bm{\beta }}}$ and ${{\cal{G}}_{\bm{\delta }}}$ represent the two decoding networks for reconstructing the semantic information and source image, respectively, with ${\bm{\beta}}$ and ${\bm{\delta}}$ denoting the corresponding trainable parameters. Here, ${\bf{\hat S}}$ and ${\bf{\hat D}}$ denote the recovered semantic information and source image.

According to \cite{Meng2025DataIF}, the MIMO-OTFS channel can be modeled as the superposition of $L$ distinct propagation paths, each of which is characterized by its delay, Doppler, angle-of-departure (AoD) and angle-of-arrival (AoA)
\begin{equation}\label{}
  {{\bf{H}}_p} = \sum\limits_{i = 1}^L {{\alpha _i}\left( {{{\bf{a}}_{\rm{R}}}\left( {{\varphi _i}} \right){\bf{a}}_{\rm{T}}^H\left( {{\phi _i}} \right)} \right)}  \otimes \left( {{{\bf{\Pi }}^{{l_i}}}{{\bf{\Delta }}^{{k_i}}}} \right) ,
\end{equation}
where $\alpha_i$, $l_i$ and $k_i$ denote the complex gain, delay tap and Doppler tap of the $i$-th path, respectively; 
${\phi _i}$ and ${\varphi _i}$ are the AoD and AoA. ${\bf{a}}_{\rm{T}}\left( {{\phi _i}} \right)$ and ${{\bf{a}}_{\rm{R}}}\left( {{\varphi _i}} \right)$ represent the array response vectors related to the corresponding angles, whose expression are provided in \cite{Zhang2022TreeCA}. ${\bf{\Pi }}$ is an $MN \times MN$ forward cyclic shift matrix, and ${\bf{\Delta }} = {\rm{diag}}\left\{ {{e^{j\frac{{2\pi q}}{{MN}}}}} \right\}_{q = 0}^{MN - 1}$ is an $MN \times MN$ diagonal phase-rotation matrix.

\section{Semantic-Aware Sub-Channel Allocation}
This section presents the core methodology for SA-SCA in the proposed MIMO-OTFS-SC system. The framework consists of three key components: (a) SVD-based precoding and combining that decomposes the MIMO-OTFS channel into parallel sub-channels characterized by singular values; (b) An EM-based semantic importance evaluation mechanism that quantifies the criticality of semantic features; and (c) A two-stage training strategy with specialized loss functions that jointly optimize semantic preservation and sub-channel matching. The fundamental insight is to maximize the alignment between semantic importance and sub-channel quality through Kendall correlation optimization, thereby ensuring that semantically critical information is preferentially allocated to higher-quality sub-channels for enhanced transmission reliability.
\subsection{SVD-Based Precoding and Combining}
Assuming the channel remains constant over $P$ OTFS frames, we denote the channel matrix uniformly as $\bf{H}$, omitting the frame-specific subscript $p$. The same precoder $\bf{G}$ and combiner $\bf{W}$ are applied to all frames. It is assumed that the RX can accurately acquire $\bf{H}$ via the channel estimation method described in \cite{Meng2025DataIF}. By performing singular value decomposition (SVD) on ${\bf{H}}$, we obtain
\begin{equation}\label{}
  {{\bf{H}}} = {{\bf{U}}}{{\bf{\Sigma }}}{\bf{V}}^H ,
\end{equation}
where ${{\bf{V}}}$ and ${{\bf{U}}}$ are $N_{\rm{T}}MN \times r$ and $N_{\rm{R}}MN \times r$ unitary matrices, respectively, satisfying ${\bf{V}}^H{{\bf{V}}} = {\bf{U}}^H{{\bf{U}}} = {{\bf{I}}_{r}}$, with $r$ being the rank of ${\bf{H}}$. Here, ${{\bf{\Sigma }}} = {\rm{diag}}\left\{ {{\lambda _{s}}} \right\}_{s = 1}^{{r}}$ is a $r \times r$ diagonal matric comprising $r$ singular values arranged in descending order.

The optimal precoder and combiner are obtained by extracting the first $N_{\rm{RF}}MN$ columns of ${\bf{V}}$ and ${\bf{U}}$, respectively. Without loss of generality, we assume $N_{\rm{RF}}MN$ is equal to $r$. Thus, the optimal precoder and combiner can be written as
\begin{equation}\label{equ:optimal_precoder}
  {\bf{G}}^{{\rm{opt}}} = {{\bf{V}}},\; {\bf{W}}^{{\rm{opt}}} = {{\bf{U}}} .
\end{equation}

Substituting \eqref{equ:optimal_precoder} into \eqref{equ:input_output} yields
\begin{equation}\label{}
  \begin{array}{l}
{{{\bf{\hat x}}}_p} = \left( {{{\bf{I}}_{{N_{{\rm{RF}}}}}} \otimes \left( {{{\bf{F}}_N} \otimes {{\bf{I}}_M}} \right)} \right){\bf{U}}^H\left( {{{\bf{U}}}{{\bf{\Sigma }}}{\bf{V}}^H} \right){{\bf{V}}}\\
 \times \left( {{{\bf{I}}_{{N_{{\rm{RF}}}}}} \otimes \left( {{\bf{F}}_N^H \otimes {{\bf{I}}_M}} \right)} \right){{\bf{x}}_p} + {{{\bf{\bar n}}}_p} = {{\bf{\Sigma }}}{{\bf{x}}_p} + {{{\bf{\bar n}}}_p} ,
\end{array}
\end{equation}
where ${{{\bf{\bar n}}}_p}={\left( {{{\bf{{I}}}_{{N_{{\rm{RF}}}}}} \otimes \left( {{{\bf{F}}_N} \otimes {{\bf{{I}}}_M}} \right)} \right){\bf{W}}^H{{\bf{n}}_p}}$ denotes the effective noise in the DD-domain.

For the $s$-th element of ${{{\bf{\hat x}}}_p}$, we have
\begin{equation}\label{}
  {\hat x_{p,s}} = {\lambda _{s}}{x_{p,s}} + {\bar n_{p,s}} ,
\end{equation}
where ${x_{p,s}}$ and ${\bar n_{p,s}}$ represent the $s$-th element of ${{\bf{x}}_{p}}$ and ${{\bf{\bar n}}_{p}}$, respectively.

Therefore, by exploiting SVD-based precoding and combining, the MIMO-OTFS channel is equivalent to $N_{\rm{RF}}MN$ independent sub-channels, and the quality of each sub-channel is characterized by its corresponding singular value.

\subsection{EM-Based Semantic Importance Evaluation}
To enhance system performance, it is essential to prioritize the assignment of more critical semantic information to sub-channels with superior quality. This strategy helps minimize information loss and improves overall transmission efficiency. The effectiveness of such sub-channel allocation heavily depends on the chosen evaluation criteria. Thus, a key challenge lies in assessing both the channel conditions of the sub-channels and the importance of the semantic information.

The sub-channel gain vector ${\bm{\gamma}}$ serves as a quantitative measure of sub-channel quality, defined as the vector composed of the singular values of all sub-channels
\begin{equation}\label{}
  {\bm{\gamma}} = {\left[ {{\lambda _1},{\lambda _2},...,{\lambda _{{N_{{\rm{RF}}}}MN}}} \right]^T} .
\end{equation}

Inspired by \cite{Wang2024WirelessAI}, we employ an entropy model (EM) to assess the importance of the semantic feature $\mathbf{f}$ output by the semantic encoder. The process begins by feeding $\mathbf{f}$ into a hyper encoder and a quantizer, producing a deeper feature $\mathbf{z}$ and a quantized representation $\hat{\mathbf{y}}$, respectively. The deeper feature $\mathbf{z}$ is then quantized to yield $\hat{\mathbf{z}}$, which is subsequently fed into a hyper decoder to obtain $\tilde{\mathbf{z}}$. 
The outputs $\hat{\mathbf{y}}$ and $\tilde{\mathbf{z}}$ are then jointly input to a Gaussian parameter estimator. The resulting estimated Gaussian parameters are passed to an entropy calculation module, producing the entropy values of the semantic feature. These entropy values serve as the semantic importance vector $\mathbf{w} = [w_1,w_2,...,w_{N_{\rm{RF}}MN}]^T$.

To measure the degree of matching between semantic importance and sub-channel conditions, we employ the Kendall correlation coefficient between the semantic importance vector $\mathbf{w}$ and the sub-channel gain vector $\bm{\gamma}$ as the evaluation metric such that
\begin{equation}\label{}
  \kappa  = \frac{{2\sum\limits_{s < s'} {{\rm{Sigmoid}}\left[ { - 2\left( {{w_s} - {w_{s'}}} \right)\left( {{\lambda _s} - {\lambda _{s'}}} \right)} \right]} }}{{{N_{{\rm{RF}}}}MN\left( {{N_{{\rm{RF}}}}MN - 1} \right)}} ,
\end{equation}
which reflects the ordinal similarity between the two sequences.

\subsection{Training and Loss Function}
The proposed model adopts a two-stage training strategy due to the functional separation between semantic importance assessment and semantic encoding/decoding. Stage I jointly trains the semantic encoder and semantic importance assessment model, with the goal of improving the latter's ability to evaluate semantic importance. Stage II fine-tunes both the semantic encoder and decoder to enhance their ability to extract semantic information and perform effective sub-channel allocation for transmission.

The first training stage aims to jointly minimize the transmitted bit-stream length and the reconstruction distortion. Since both the latent feature ${\bf{\hat{y}}}$ and hyper-latent feature ${\bf{\hat{z}}}$ contribute to the transmitted bit-stream, the loss function must account for the transmission cost of ${\bf{\hat{y}}}$ and ${\bf{\hat{z}}}$. The first-stage loss function, incorporating mean square error (MSE) as the distortion measure, is formulated as
\begin{equation}\label{}
  {{\mathcal{L}}_1} = {\mathbb{E}}\left[ { - {{\log }_2}{p_{{\bf{\hat y}}}}\left( {{\bf{\hat y}}} \right)} \right] + {\mathbb{E}}\left[ { - {{\log }_2}{p_{{\bf{\hat z}}}}\left( {{\bf{\hat z}}} \right)} \right] + \left\| {{\bf{D}} - {\bf{\hat D}}} \right\|_2^2 ,
\end{equation}
where the first two items correspond to the bit-stream lengths of ${\bf{\hat{y}}}$ and ${\bf{\hat{z}}}$, respectively, while the third item evaluates reconstruction fidelity by comparing the output image with its original counterpart.

During the second training phase, the goal is to maximize recovery of both source data and semantic content from the received feature sequence $\{{\bf{\hat x}}_n\}_{n=1}^{N_{\rm{Sa}}}$, while ensuring that important semantic information is matched with high-quality sub-channels. The loss function is defined as
\begin{equation}\label{equ:loss2}
  \begin{array}{l}
{{\mathcal{L}}_2} =  - \frac{1}{{{N_{{\rm{Sa}}}}}}\sum\limits_n^{{N_{{\rm{Sa}}}}} {\sum\limits_l^{{N_{{\rm{La}}}}} {p\left( {{S_n} = l|{{{\bf{\hat x}}}_n}} \right)\log {p_{{{\mathcal G}_\beta }}}\left( {{S_n} = l|{{{\bf{\hat x}}}_n}} \right)} } \\
 + {\lambda _{{\rm{MSE}}}}\left\| {{\bf{D}} - {\bf{\hat D}}} \right\|_2^2 - {\lambda _{{\rm{EM}}}} \cdot \kappa \;,
\end{array}
\end{equation}
where $N_{\rm{Sa}}$ denotes the number of samples per training batch. In the context of the image classification task considered here, the semantic information is confined to classification labels, i.e. ${S_n} \in \left\{ {1,2,...,{N_{{\rm{La}}}}} \right\},n = 1,2,...,{N_{{\rm{Sa}}}}$, where $S_n$ represents the semantic label of the $n$-th sample, and $N_{\rm{La}}$ is the number of classes. The cross-entropy (CE) loss, appearing as the first term in \eqref{equ:loss2}, measures the mismatch between the ground-truth and reconstructed classification distributions, with ${{p_{{{\mathcal G}_\beta }}}\left( {{S_n} = l|{{{\bf{\hat x}}}_n}} \right)}$ denoting the classification probability output by the semantic decoder ${{\mathcal G}_\beta }$. The coefficients $\lambda _{\rm{MSE}}$ and $\lambda _{\rm{EM}}$ are the trade-off parameters that balance the contributions of image classification accuracy, reconstruction quality, and sub-channel matching.

\section{Simulation Results} \label{sec-sim}

\begin{figure} [bt!]
	\centering
	\includegraphics[width=3.2in]{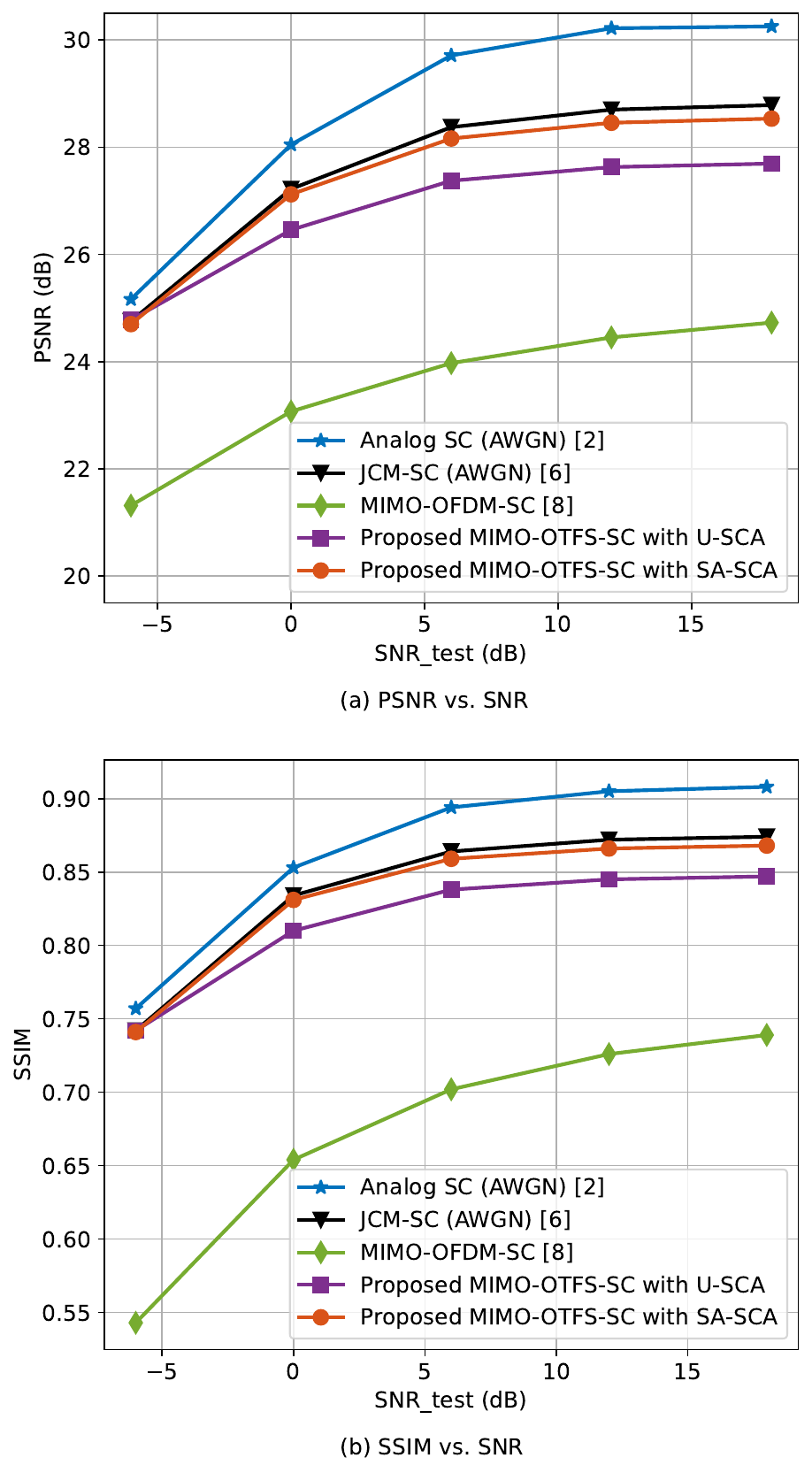}
	\caption{System performance vs. SNR with $N_{\rm{T}}=N_{\rm{R}}=8$ and $N_{\rm{RF}}=2$.}
	\label{fig:sim_SNR_PIMRC}
\end{figure}

\begin{figure} [bt!]
	\centering
	\includegraphics[width=3.2in]{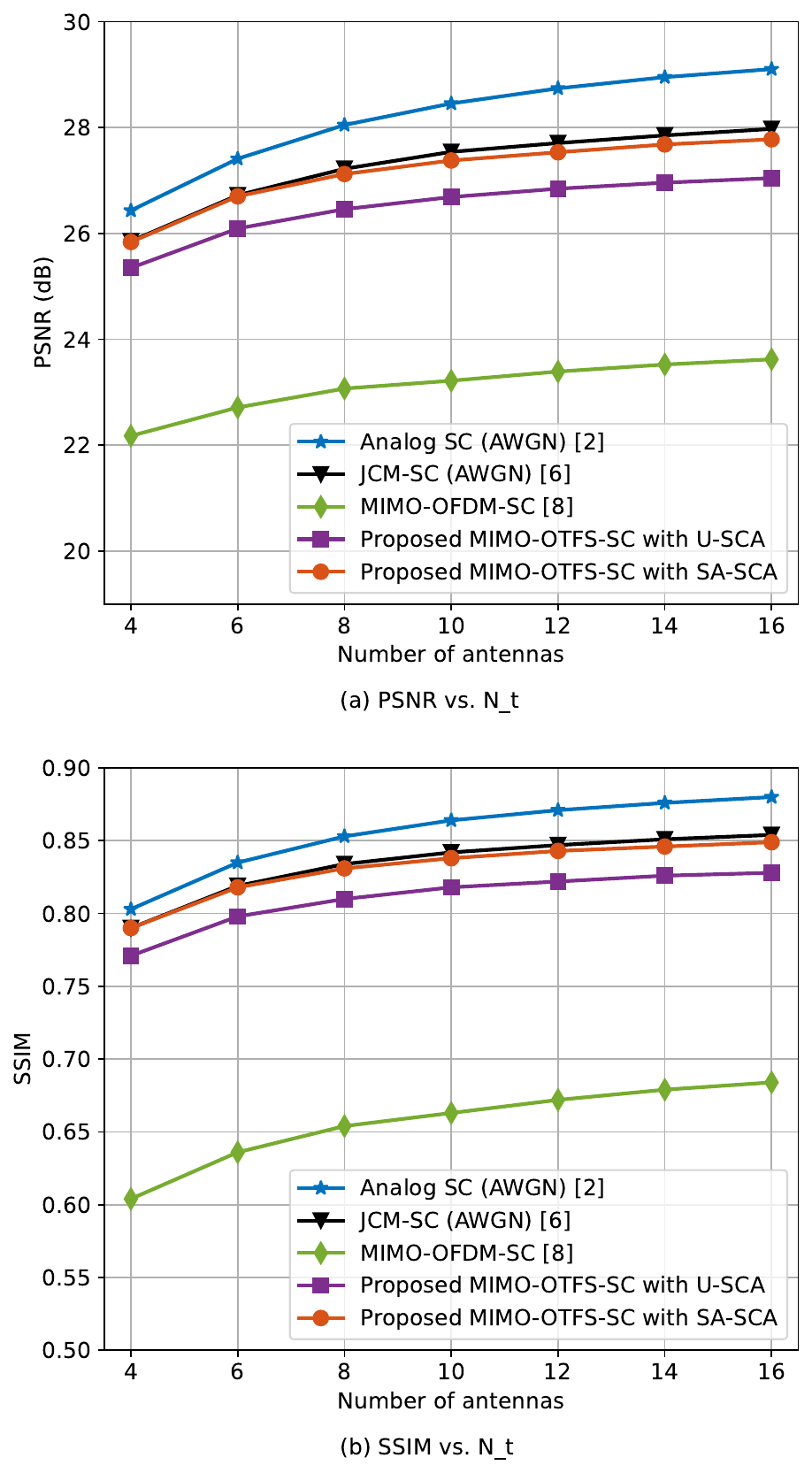}
	\caption{System performance vs. $N_{\rm{T}}$ with $N_{\rm{RF}}=2$ and ${\rm{SNR}}_{\rm{test}}=0$.}
	\label{fig:sim_Nt_PIMRC}
\end{figure}

We assess the proposed semantic-aware MIMO-OTFS-SC system by performing image transmission over a high-mobility channel environment with the maximum velocity of 500km/h. The system is equipped with $N_{\rm{RF}}=2$ RF chains, utilizing 64-QAM constellation modulation. The center frequency is configured as 28 GHz and the subcarrier spacing as 120 kHz. The OTFS parameters are configured with $M=8$ (delay dimension) and $N=8$ (Doppler dimension). The channel comprises $L=10$ independent paths, with maximum delay and Doppler taps of 5 and 1, respectively. The delay taps $l_i$ and Doppler taps $k_i$ are randomly generated. The gain of each path $\alpha_i \sim {\cal C}{\cal N}\left( {0,1} \right)$. The azimuth angles ${{\varphi _i}}$ and ${{\phi _i}}$ are uniformly distributed in $[0,\pi]$.

The neural network architectures of semantic encoder, probability converter, semantic information reconstruction, and source image reconstruction modules follow the design in \cite{Bo2024JointCM}. The channel use is set to $1024$. For both training stages, we use a batch size of $32$ and optimize the model using the Adam optimizer with a cosine annealing schedule, where the learning rate decays from an initial value of $5 \times 10^{-4}$ to $10^{-6}$ over $300$ epochs. The EM structure is adopted from \cite{Wang2024WirelessAI} and is trained separately with the Adam optimizer. The learning rate is set to a constant value of $10^{-4}$. The trade-off parameters in the loss function are empirically set as $\lambda_{\rm{MSE}}=20$ and $\lambda_{\rm{EM}}=0.5$.

The proposed MIMO-OTFS-SC with SA-SCA is compared against analog semantic communication \cite{Bourtsoulatze2019DeepJS} under AWGN channel, JCM-based semantic communication (JCM-SC) \cite{Bo2024JointCM} under AWGN channel, MIMO-OFDM-based semantic communication (MIMO-OFDM-SC) \cite{Yang2022OFDMGD} under doubly-selective channel, and MIMO-OTFS-SC with uniform sub-channel allocation (U-SCA). All experiments are conducted using the CIFAR-10 dataset. The model is implemented in PyTorch 1.13.1 with CUDA 11.7 and executed on an NVIDIA A800 GPU with 80 GB memory. 
%

\begin{table}[bt!]
	\centering
	\renewcommand{\arraystretch}{1.2}
	\caption{Classification accuracy of various semantic communication schemes under different SNR test conditions}
	\label{table:accuracy}
	\begin{tabular}{|c|c|c|c|c|c|}
		\hline
		{\bf Algorithm / SNR} & -6 dB & 0 dB & 6 dB & 12 dB & 18 dB\\
		\hline
		Analog SC & \textbf{0.879} & 0.867 & 0.869 & 0.870 & 0.870 \\
		\hline
        JCM-SC & 0.874 & 0.873 & 0.875 & 0.876 & 0.875 \\
		\hline
        MIMO-OFDM-SC & 0.873 & 0.871 & 0.873 & 0.876 & 0.875 \\
		\hline
		\makecell{MIMO-OTFS-SC \\ with U-SCA} & 0.877 & 0.869 & 0.869 & 0.868 & 0.868 \\
		\hline
        \makecell{MIMO-OTFS-SC \\ with SA-SCA} & 0.876 & \textbf{0.878} & \textbf{0.877} & \textbf{0.878} & \textbf{0.878} \\
		\hline
	\end{tabular}
\end{table}

Ideal CSI is assumed known to the transceiver. In the training stage, the number of antennas is set to $N_{\rm{T}}=N_{\rm{R}}=8$, and the SNR ($\mathrm{SNR}_{\mathrm{train}}$) is set to 0 dB. Then the trained model is evaluated at SNRs of -6, 0, 6, 12, and 18 dB. Table \ref{table:accuracy} lists the classification accuracy of various semantic communication schemes under different SNR test conditions. When the SNR is greater than 0 dB, the proposed MIMO-OTFS-SC with SA-SCA outperforms all competing schemes in classification accuracy.

Fig. \ref{fig:sim_SNR_PIMRC} presents the peak signal-to-noise ratio (PSNR) and structural similarity index measure (SSIM) performance of various semantic communication schemes. As the SNR increases, both the PSNR and SSIM of all schemes show an improving trend. However, under the same SNR condition, the image quality performance of the MIMO-OFDM-SC scheme consistently lags behind the other schemes. For instance, the PSNR of MIMO-OFDM-SC is 3.38 dB and 4.05 dB lower than that of MIMO-OTFS-SC with U-SCA and SA-SCA, respectively. Within the proposed MIMO-OTFS-SC scheme, the SA-SCA mechanism outperforms the U-SCA approach, achieving reconstruction quality close to that of the JCM scheme under AWGN channels, which validates the effectiveness of the proposed method in time-frequency doubly-selective channels. In summary, the proposed MIMO-OTFS-SC scheme, particularly with the SA-SCA strategy, maintains high classification accuracy and superior image reconstruction quality, demonstrating its robustness and practical potential in challenging channel environments.

Then the trained model is evaluated with numbers of transmit antennas $N_{\rm{T}} = 4, 6, 8, 10, 12, 14, 16$, while keeping $N_{\rm{R}}=N_{\rm{T}}$ and $N_{\rm{RF}}=2$. Fig. \ref{fig:sim_Nt_PIMRC} shows the PSNR and SSIM performance as functions of the number of antennas. Both PSNR and SSIM gradually improve as $N_{\rm{T}}$ increases. Among all schemes, MIMO-OFDM-SC shows the lowest PSNR and SSIM, further highlighting its limitations in signal recovery over doubly-selective channels. In contrast, the proposed MIMO-OTFS-SC scheme with SA-SCA achieves a PSNR of up to approximately 28 dB and an SSIM of 0.85 at $N_{\rm{T}} = 16$, significantly outperforming MIMO-OFDM-SC. These results confirm that increasing the number of antennas enhances overall system performance, and the proposed MIMO-OTFS-SC scheme, particularly with SA-SCA, maintains  superior image reconstruction quality, demonstrating its robustness and practical potential in challenging channel environments.

\section{Conclusion}
This paper presents a semantic communication framework integrating MIMO-OTFS with semantic-aware resource allocation for robust image transmission in high-mobility environments. Our key contributions include a novel MIMO-OTFS architecture resilient to doubly-selective fading and a semantic importance-aware sub-channel allocation strategy that optimally maps critical features to quality sub-channels via entropy modeling and Kendall correlation. Experimental results demonstrate significant gains in reconstruction quality over conventional OFDM-based semantic communication system under doubly-selective channels, validating the effectiveness of semantic-physical layer co-design.

\section*{Acknowledgement}

This work was supported in part by the National Natural Science Foundation of China under Grant 62293484, in part by the National Key Research and Development Program of China under Grant 2023YFB2904300, and in part by the Program of Jiangsu Province under Grant NTACT-2024Z-001.

\end{document}